\documentclass[twocolumn,showpacs,preprintnumbers,amsmath,amssymb,prb]{revtex4}

\usepackage{graphicx}
\usepackage{dcolumn}
\usepackage{bm}

\begin{document}

\title{Nonadiabatic Time-Dependent Spin-Density Functional Theory for strongly correlated systems}
\author{ Volodymyr~Turkowski\altaffiliation{Corresponding author, e-mail address: Volodymyr.Turkowski@ucf.edu} and 
Talat S.~Rahman}

\affiliation{Department of Physics and NanoScience and Technology Center,
University of Central Florida, Orlando, FL 32816}

\date{\today}

\begin{abstract}
We propose a nonadiabatic time-dependent spin-density functional theory (TDSDFT) approach for studying single-electron excited states and ultrafast response of systems with strong electron correlations. The correlation part of the nonadiabatic exchange-correlation (XC) kernel is constructed by using exact results for the Hubbard model of strongly correlated electrons. We demonstrate that the corresponding nonadiabatic XC kernel reproduces main features of the spectrum of the Hubbard dimer and the 2D, 3D and infinite-dimensional Hubbard models, some of which are impossible to obtain within the adiabatic approach. The formalism may be applied for {\it ab initio} examination of strongly correlated electron systems in- and out-of-equilibrium within the TDSDFT, extending it beyond the metallic and semiconductor structures with plasmons, excitons and other excitations.
\end{abstract}


\maketitle

{\it Introduction}.--Reliable description of the electronic properties of systems that contain localized d- and f-orbitals remains a challenging problem in condensed- matter physics for both extended and finite systems. Extended systems form a large class of materials with many exotic properties which in turn lead to many potential technological applications. Examples include cuprate high-temperature superconductors, heavy fermion materials and manganites. One may also expect unusual properties in the much less explored case of molecules and nanostructures, in which correlation effects may be  even more enhanced owing to space confinement.  Moreover, in nanotechnological applications, in which the distance between the atoms can be tuned (for example, by putting the atoms on a substrate), even s- and p-electron systems may find themselves in a strongly-correlated regime when the interatomic separation is appropriately large. These nanoscale systems show many surprises, such as the recently observed unusual antiferromagnetic ordering in small Fe chains,\cite{1} and the metal-insulator transition in Au and Fe chains.\cite{2} The excited states and the nonequilibrium properties of such systems, including their ultrafast (femto- and atto-second) response, are very relevant to the modern trend of designing "smaller and faster" systems. Correct understanding of nanosystems and molecules with strong electron correlations may also shed light on the general properties of strongly correlated materials, including local correlations and nonhomogeneous order in extended systems. 

Most of the progress in studies of correlated systems has been made by using many-body approaches. Two of the most powerful are the Bethe ansatz for 1D systems and dynamical mean-field theory (DMFT), which is exact in the limit of infinite dimensions\cite{3} and appears to be also a good approximation for 2D and 3D systems (reviews of both approaches are given in Refs.~\onlinecite{4} and \onlinecite{5}, respectively). These methods have also been generalized for the nonequilibrium case,\cite{6,7} allowing one to study the excitations and the nonlinear response of corresponding systems. Since standard DFT approximations fail to describe properly most strongly-correlated effects, 
DMFT combined with DFT has been proposed as an alternative.\cite{8,9} 
The latter allows one to describe spectral, optical and magnetic properties of bulk and layered materials  (see, for example, reviews in Refs. \onlinecite{10,11}). In DFT+DMFT, all properties of "non-correlated" systems (system geometry, bandstructure, etc.) are obtained within DFT (usually with the LDA or the GGA approximation), and correlation effects are taken into account by solving the corresponding effective Hubbard model. Recently, it has been shown that a similar approach can be successfully applied to nanosystems.\cite{12,13,14,15,16,17,18}  Still, the combined DFT+DMFT calculations are computationally
demanding even for the equilibrium case, especially for finite systems, which have a number of non-equivalent atoms. 

For this reason it would be very useful to develop a TDDFT formalism with XC potential that properly describes strongly correlated systems, including their excitations (in which case one needs to go beyond static DFT).\cite{19} Some progress in this direction has already been made. For 1D systems, appropriate strongly correlated adiabatic (static DFT) XC potentials were proposed in Refs.~\onlinecite{20,21,22,23}. In particular, in Refs.~\onlinecite{22,23} a DFT approach based on the Bethe ansatz has been proposed. Studies of the Kondo effect in the Anderson impurity model with DFT \cite{Stefanucci,Bergfield} and the Hubbard model with lattice DFT\cite{Pastor} have also been presented.
Recently, a combined adiabatic TDDFT-DMFT method applicable for the 3D case has been proposed,\cite{24}
in which the XC potential is derived from the DMFT solution for the Hubbard model. The above approaches were tested 
through comparison of some results with exact solutions, and it was shown that they 
successfully reproduce several important effects, including metal-insulator transition and temporal response of systems. 
There have been some drawbacks too.  In particular, as was shown in Ref.~\onlinecite{24}, the proposed adiabatic potential fails to describe correctly the response of a finite Hubbard system in the strongly correlated regime when the system is close to half-filling and when the local Coulomb repulsion U is large. One might expect that the main shortcoming comes from the adiabatic approximation. Indeed, as one notes, the adiabatic XC kernel fails to even reproduce the correct number of excited states of the system. For example, in the case of the Hubbard dimer, it is known that the local Coulomb repulsion -- in addition to the bonding-anti-bonding orbital excitation -- leads to a satellite level at finite U (see, e.g., Ref.~\onlinecite{25}). This extra state is related to the excited Hubbard "band" in the case of extended systems (in fact it may be traced to the single-electron spectral weight to a higher energy ($\sim$ U)). It is also known from the DMFT solution that Hubbard systems demonstrate an extra quasi-particle spectral weight at zero energy (chemical potential).\cite{5} Although all of these states are important in a strongly correlated regime, none can be reproduced with the adiabatic TDDFT. Indeed, in the adiabatic case the solution of the Casida equation\cite{26} leads to a shift of the single-electron levels, and not to new states. 

In this letter, we propose –-- on the basis of some exact results for the Hubbard model –-- a simple form of the nonadiabatic XC kernel which results in the single-electron spectrum of the Hubbard model that reproduces the main features of the spectra of both finite (dimer) and extended (infinite-dimensional) systems. This kernel can be easily implemented within the standard TDDFT codes for use for strongly correlated systems.

{\it The Hubbard dimer}.--In order to obtain the XC kernel $f_{XC\sigma\sigma '}({\bf r}, {\bf r}',\omega)$ 
for the Hubbard dimer, we map  the eigenvalue equation (which defines the positions of the spectral peaks) derived from the dimer Green's function onto
the corresponding TDDFT Casida eigenenergy equation,\cite{26} which has the following general form:
\begin{equation}
{\rm det} \left[
\begin{array}{c}
\epsilon_{0}^{2}+2\epsilon_{0}K_{\uparrow ,\uparrow}(\omega ) -\omega^{2}\\
2\epsilon_{0}K_{\downarrow ,\uparrow}(\omega )
\end{array}
\begin{array}{c}
2\epsilon_{0}K_{\uparrow ,\downarrow}(\omega )\\
\epsilon_{0}^{2}+2\epsilon_{0}K_{\downarrow ,\downarrow}(\omega) -\omega^{2}
\end{array}
\right] =0,
\label{Casida}
\end{equation}
where 
\begin{eqnarray}
K_{\sigma ,\sigma '}(\omega )=\int\int \psi_{g}({\bf r})\psi_{u}({\bf r})\left(\frac{1}{|{\bf r}-{\bf r}'|}
\right.
\nonumber \\
\left.
+f_{XC\sigma, \sigma '}(\bf r,{\bf r}',\omega)\right)\psi_{g}({\bf r}')\psi_{u}({\bf r}') 
d{\bf r}d{\bf r}'
\label{K}
\end{eqnarray}
(the first part being the Hartree term, and $f_{XC\sigma, \sigma '}(\bf r,{\bf r}',\omega)$ the Fourier transform of $\delta V_{XC\sigma}[n]({\bf r},t)/\delta n_{\sigma '}({\bf r}',t')$ with respect to t-t'), 
$\epsilon_{0}=\epsilon_{u}-\epsilon_{g}$ is the excitation energy of the free electron,
$\epsilon_{u,g}$ and $\psi_{u,g}({\bf r})$ are the corresponding bonding- and anti-bonding energies and wave functions.
This equation has two solutions
\begin{eqnarray}
\omega^{2}=\epsilon_{0}(\epsilon_{0}+2(K_{\uparrow ,\uparrow}\pm K_{\uparrow ,\downarrow})),
\label{Casidasolution}
\end{eqnarray}
where $+$ corresponds to the singlet state, and $-$ to the triplet one. Since the ground state of this system is a singlet
and the total spin of the isolated system is conserved, for definiteness we shall focus on the singlet state. 

The dimer Green's function can be found from the exact solution for the Hubbard model with the Hamiltonian:
\begin{eqnarray}
H=-t\sum_{i\not=j,\sigma}c_{i\sigma}^{\dagger}c_{j\sigma}+U\sum_{i}n_{i\uparrow}n_{i\downarrow},
\label{Hubbard}
 \end{eqnarray}
where $c_{i\sigma}^{\dagger}$ and $c_{j\sigma}$ are the creation and annihilation operator of electron with spin $\sigma$ on site i, $n_{i\sigma}$ is the corresponding number operator, and t is the hopping parameter.
The exact single particle dimer Green's function
has the following form in the singlet (or, more generally, the "non-magnetic") case:
\begin{equation}
{\hat G}^{-1}(\omega )=
\left( \begin{array}{c}
\omega-\Sigma_{11}(\omega )\\
t-\Sigma_{21}(\omega )
\end{array}
\begin{array}{c}
t-\Sigma_{12}(\omega )\\
\omega-\Sigma_{22}(\omega )
\end{array}
\right) ,
\label{G}
\end{equation}
where the self-energies are
\begin{eqnarray}
\Sigma_{11,\sigma\sigma '}=\Sigma_{22,\sigma\sigma '}=\delta_{\sigma\sigma '}\frac{U^{2}}{8}\left( 
\frac{1}{\omega -3t+i\delta}+\frac{1}{\omega +3t+i\delta}
\right),\nonumber \\
\label{Sigma11}
\end{eqnarray}
and
\begin{eqnarray}
\Sigma_{12,\sigma\sigma '}=\Sigma_{21,\sigma\sigma '}=\delta_{\sigma\sigma '}\frac{U^{2}}{8}\left( 
\frac{1}{\omega -3t+i\delta}-\frac{1}{\omega +3t+i\delta}
\right).\nonumber \\
\label{Sigma12}
\end{eqnarray}
Substitution of Eqs.~(\ref{Sigma11}) and (\ref{Sigma12}) into Eq.~(\ref{G}) leads to the following
eigenvalue equation:
\begin{equation}
t^{2}+\frac{U^{2}}{2}+\frac{U^{2}}{2}\frac{6t^{2}-U^{2}/4}{\omega^{2}-9t^{2}}
=\omega^{2}.
\label{eigenproblem}
\end{equation}
From Eqs.~(\ref{Casidasolution}) and (\ref{eigenproblem}), one can then obtain an equation for the XC kernel for the Hubbard dimer:
\begin{eqnarray}
K_{\uparrow\uparrow}(\omega )+K_{\uparrow\downarrow}(\omega )=\frac{U^{2}}{4t}\left(1+\frac{6t^{2}-U^{2}/4}{\omega^{2}-9t^{2}}\right) ,
\label{Kdimer}
\end{eqnarray}
where $K_{\sigma\sigma '}(\omega )$ is 
 in Eq.~(\ref{K}).
Since we are interested in the contribution to the energy from correlation effects,
we assume that the Hartree ($\sim 1/|{\bf r}-{\bf r}'|$) and the exchange ($K_{\uparrow\uparrow}$)
parts simply lead to the renormalization of the free-particle energy $\epsilon_{0}\rightarrow t$ in Eq.~(\ref{Casida}).

Farthermore, from Eqs.(\ref{K}) and (\ref{Kdimer}), one can arrive at the following separable form of the correlation portion of the local XC kernel:
\begin{eqnarray}
f_{C\uparrow, \downarrow}({\bf r},{\bf r}',\omega)
=\delta ({\bf r} - {\bf r}')
\frac{U^{2}}{4tA}\left(1+\frac{6t^{2}-U^{2}/4}{\omega^{2}-9t^{2}}\right),
\label{fXCdimer}
\end{eqnarray}
were $A=\int\int |\psi_{g}|^{2}({\bf r})|\psi_{u}|^{2}({\bf r})d{\bf r}d{\bf r}'$.
This kernel results in the exact excitation spectrum (or more precisely, positions of the spectral function peaks) of the Hubbard dimer\cite{25}:
\begin{eqnarray}
\omega=\pm t\pm\sqrt{4t^{2}+\frac{U^{2}}{4}}.
\label{Edimer}
\end{eqnarray}
The spectra obtained from Eq.~(\ref{Edimer}), presented in Fig.1a, show two extra states: Hubbard satellite peaks with energy $\pm 3t$ at small Us. These states appear only at finite U and arise from the redistribution of the single-electron spectral weight. It is easy to see that one cannot obtain these new states from Eq.~(\ref{Casida}) with the static XC kernel.

In the case of extended systems, considered below, the DMFT approximation 
$\Sigma_{ij}(\omega)\simeq \delta_{ij}\Sigma (\omega) $ is valid. Even though in the case of the dimer this approximation is not sufficiently accurate, we present the corresponding expression for the kernel for the discussion below:
\begin{eqnarray}
f_{C\uparrow, \downarrow}(\bf r,{\bf r}',\omega)&=&\delta ({\bf r} - {\bf r}')
\frac{U^{2}}{32tA}\left(8+\frac{72t^{2}-U^{2}}{\omega^{2}-9t^{2}}
\right.
\nonumber \\
&~&\left.-\frac{9t^{2}U^{2}}{(\omega^{2}-9t^{2})^{2}}
\right) .
\label{fXCdimerDMFT}
\end{eqnarray}
As follows from Eqs. (\ref{fXCdimer}) and (\ref{fXCdimerDMFT}), they have the same expression in the high-frequency limit, but as $\omega\rightarrow 0$ they differ. The DMFT solution has an extra energy peak around zero energy, which is a characteristic of this approach (Fig.1).

\begin{figure}[t]
\includegraphics[width=8.0cm]{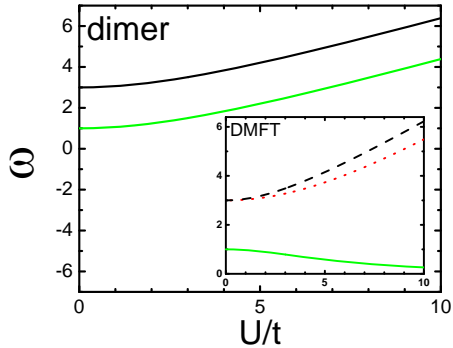}
\caption{\label{fig1} 
Exact and DMFT (insert) excitation spectra for the Hubbard dimer as a function of $U/t$. The de-excitation (negative) energies are not shown. The energy on the y-axis  is given in units of t.  
The green (solid) lines are the standard bonding-anti-bonding transitions, while the additional curves (black (dash) 
and red (dot)) correspond to the extra Hubbard (satellite) peaks which come from the correlation effects.}
\end{figure}

{\it Extended systems}.--In these cases the DMFT with the local self-energy approximation can be applied. We shall use the following exact result for the high-frequency electron self-energy for the Hubbard model:
\begin{eqnarray}
\Sigma_{ij}(\omega)=\delta_{ij}\left(    
U_{i}n_{i\downarrow}+\frac{U_{i}^{2}n_{i\downarrow}(1-n_{i\downarrow})}{\omega}
\right) .
\label{SigmaR}
\end{eqnarray}
(see, for example Refs.~\onlinecite{27,28}).
In the homogeneous case, one can construct the XC kernel by mapping the eigenvalue equation
\begin{eqnarray}
\omega -\varepsilon_{k}-\Sigma (\omega )=0,
\label{DMFT}
\end{eqnarray}
or 
\begin{eqnarray}
\varepsilon_{k}^{2}+2\varepsilon_{k}Re[\Sigma (\omega )]+|\Sigma (\omega )|^{2}=\omega^{2},
\label{DMFT}
\end{eqnarray}
onto the corresponding Casida equation.
As in the dimer case one can find the equation which connects the XC kernel with the self-energy:
\begin{eqnarray}
K_{\uparrow\downarrow} (\omega )\sim Re[\Sigma (\omega )] .
\label{KDMFT}
\end{eqnarray}
A more straightforward way to find the expression for the XC kernel is to compare the TDDFT and DMFT correlation energies:
\begin{eqnarray}
E_{C}^{TDDFT}=\frac{1}{2}\int\int \delta n({\bf r},t)f_{XC}({\bf r},t;{\bf r}',t') 
\nonumber \\
\times\delta n({\bf r}',t')d{\bf r}dtd{\bf r}'dt'
\label{TDDFTenergy}\\
E_{C}^{DMFT}=\int\int \psi^{*}({\bf r},t)\Sigma (t-t')\delta({\bf r}-{\bf r}')
\nonumber \\
\times\psi ({\bf r}',t')d{\bf r}dtd{\bf r}'dt'
\label{DMFTenergy}
\end{eqnarray}
Equations (\ref{SigmaR}), (\ref{TDDFTenergy}) and (\ref{DMFTenergy}),  together with the result for the dimer, 
Eq.~(\ref{fXCdimer}),  can be used to construct the following "universal" (DMFT) function for the correlation part of the XC kernel for the extended systems:
\begin{eqnarray}
f_{C\uparrow\downarrow}({\bf r},{\bf r}',\omega)=
\frac{U^{2}F[n_{0}]({\bf r})}{4t}\delta ({\bf r} - {\bf r}')
\frac{n_{\downarrow}(1-n_{\downarrow})\omega}{\omega^{2}-B^{2}} ,
\label{fXCDMFT}
\end{eqnarray}
where $F[n_{0}]({\bf r})$ is a functional of the ground state density $n_{0}({\bf r})$ 
(in general, of the spin parts) and $B\sim 3t$ in the case of dimer, while one can choose
$B^{2}$ to be equal to the mean square of the kinetic energy $\varepsilon_{k}^{2}$,
or more generally $B^{2}=\alpha\varepsilon_{k}^{2}$, where $\alpha\sim 1$.
Here $n_{\downarrow}$ is average number of the spin-down electrons per site. It is important to note that we do not include the contribution of the static Hartree term $Un_{\downarrow}$ to the self-energy, since it is canceled by the chemical potential at half-filling.
With this XC kernel one can reproduce the main features of the spectrum of the infinite-dimensional Hubbard model: split (by energy U) Hubbard bands and the zero energy quasi-particle peak, which disappears  as U increases (Fig.2).

\begin{figure}[t]
\includegraphics[width=8.0cm]{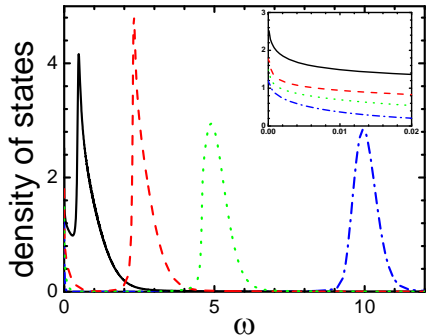} 
\caption{\label{fig2} 
a) The DOS (arbitrary units) for the infinite-dimensional Hubbard model with the hypercubic DOS 
$A(\epsilon)=(1/\sqrt{\pi}t^{*})exp (-\epsilon^{2}/t^{*2})$ ($t^{*}=2dt$ is the renormalized hopping, 
d - dimensionality of the system) and
$F[n_{0}]({\bf r})$=1, A=1, $B=\varepsilon_{{\bf k}}$. The frequency on the x-axis is given in units of $t^{*}$.
Black (solid), red (dash), blue (dot) and green (dash-dot) lines correspond to the density of states fo U=1, 5, 10 and 20,
respectively. In the insert the DOS at low frequencies is shown (due to the electron-hole symmetry
at half-filling the DOS is even function of frequency).
}
\end{figure}

\begin{figure}[t]
\includegraphics[width=8.0cm]{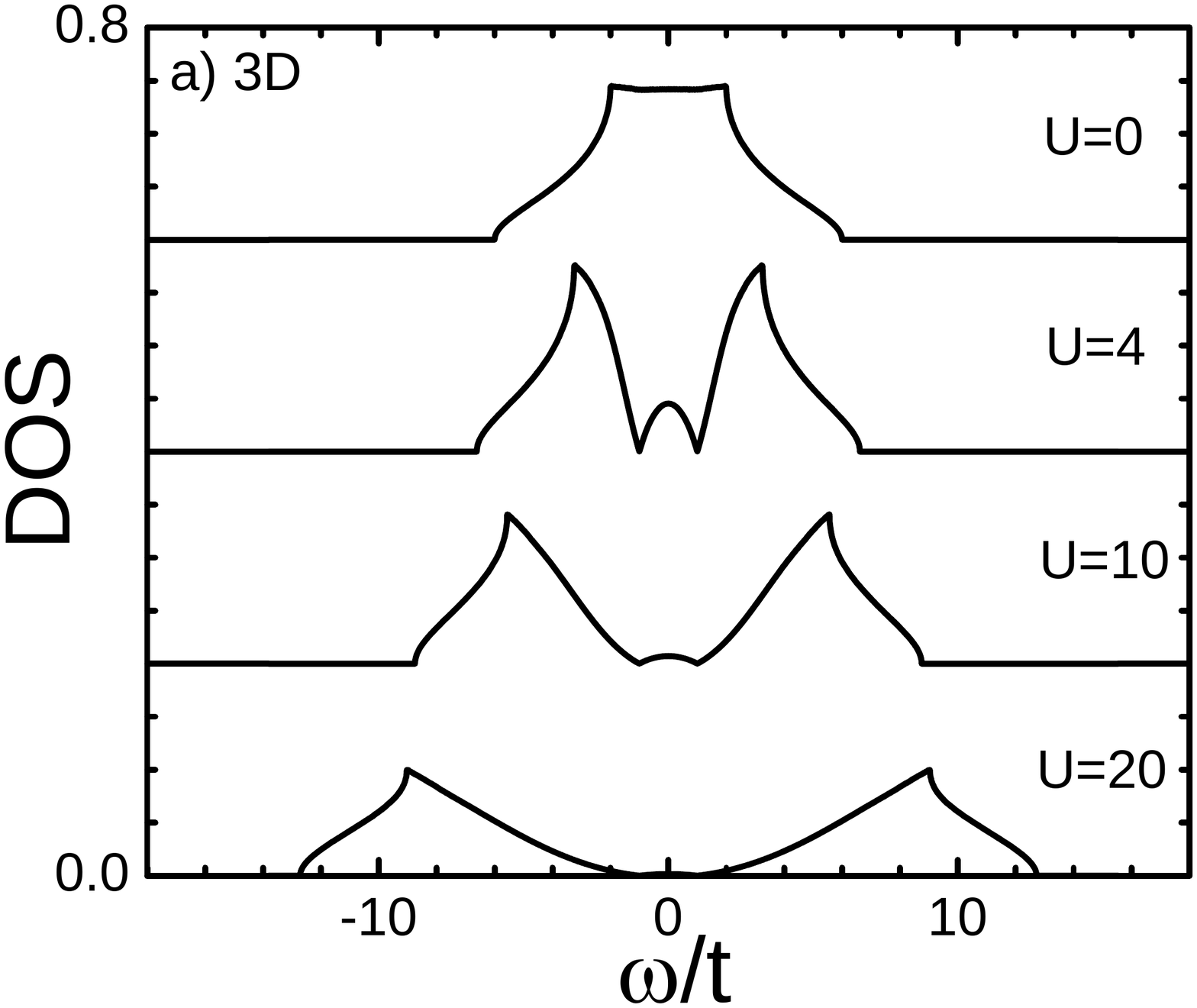}
\includegraphics[width=8.0cm]{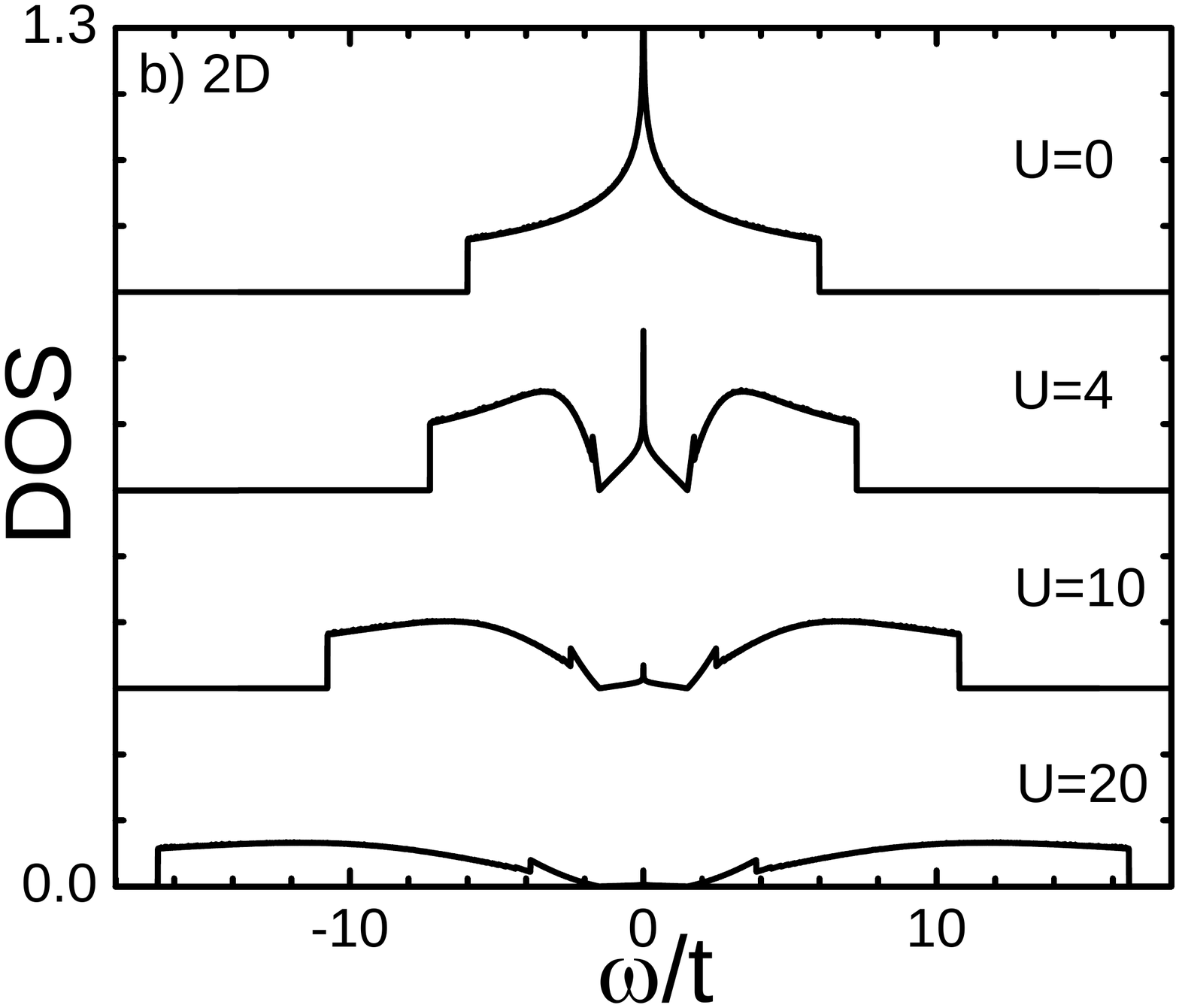}
\caption{\label{fig2} 
a) The DOS (arbitrary units) for a) the cubic lattice 3D and b) square lattice 2D Hubbard models at different values of U. The kernel parameters
are  $F[n_{0}]({\bf r})$=1, A=1, $B=t, \alpha =1$.}
\end{figure}

It is important that the above frequency dependence (nonadiabaticity) of the correlation kernel
describes the main properties of the spectrum of the dimer and the infinite-dimensional model. The spatial dependence of the kernel is not unique in our case, and can have different forms for different systems. However, as we have demonstrated above, in order to describe correctly the main features of the single-band Hubbard model (or dimer), only the frequency dependence of the kernel is essential.

In order to demonstrate that the proposed kernel can reproduce also the main features in the cases intermediate between the infinite-dimensions and dimer,
we performed calculations for the case of 3D and 2D Hubbard models. The results in Fig.3 demonstrate that the nonadiabatic kernel reproduces the Hubbard
band peaks at $\omega =\pm U/2$ and the zero energy quasi-particle peak, in agreement, for example, with DMFT calculations.\cite{Zitko,Kyung}

{\it Conclusions}.-- It is already known from the DFT+DMFT study of  real materials that the main effects of strong electron correlations can be described by local, temporally resolved electron interaction.
Here we have shown that a {\it spatially local, nonadiabatic} XC kernel is sufficient for the description of strongly correlated systems within the TDSDFT. From our analysis above it follows that one may use the following nonadiabatic XC kernel to describe the main effects of strong electron-electron correlations within TDDFT:
\begin{eqnarray}
f_{XCl\uparrow;\downarrow m}({\bf r},{\bf r}',\omega)=
\frac{U_{lm}^{2}F[n_{\sigma 0}]({\bf r})}{4t}\delta ({\bf r} - {\bf r}')
\nonumber\\
\times
\frac{n_{m\downarrow}(1-n_{m\downarrow})\omega^{\alpha}}{\omega^{2}-B_{m}^{2}} ,
\nonumber \\
\label{fXC}
\end{eqnarray}
where $l$ and $m$ are the orbital indices, $B_{m}$ is a parameter proportional to the kinetic energy of band $m$ (B=3t in the case of the dimer). The functional 
$F[n_{\sigma 0}]({\bf r})$  the "strength" of spatially local correlations, and is defined by the static ("non-correlated" DFT) spin density distributions. It must have maximum at the points of the d- and f-charge localization, i.e. in a close vicinity to the atom were the dynamical interaction takes place. Therefore, in order to model strong on-site correlations it can be chosen to be proportional to the corresponding static d- and f-electron spin density. 
On the other hand in most general case of very complex material systems (including the case of strong density fluctuations), the spatial form of the correlation kernel may be nonlocal.
The frequency power $\alpha$ in the numerator is introduced heuristically in order to obtain universal formula for small and large systems and is expected to lie between zero (for small systems, such as dimer) and 1 (for the extended (DMFT) case).

It is important to notice that Eq.~(\ref{fXC}) contains proposed generalized kernel to include (heuristically) multi-orbitals, which is more relevant for real materials. Namely, the correlation potential is sum
of the terms which correspond to the intra-orbital, $\sim U_{ll}^{2}$, and inter-orbital,
$\sim U_{lm}^{2}$ spin density interactions.  In the case of several-electrons per site, one also needs to take into 
account the Hund coupling (J-) terms, even though typically J is order of magnitude smaller than the local Coulomb repulsion U. In the last case, one can take into account the effects of J in a mean-field approximation.
In particular, the spin flip J-terms in the Hubbard model can be taken into account by using the Hartree-Fock type splitting of the four fermion-operator terms. This will lead to a renormalization of the “free electron” bandstructure. Such approximation is used in the DFT+U case. Therefore, it will correspond to “static J-dynamic U” interaction in the theory presented in this article. Moreover, the contribution of J can be taken into account through renormalized Us which correspond to the repulsion between one-orbital opposite-spin electrons and different orbitals same-spin electrons (see, e.g., Ref.~\cite{11}). 
Therefore, proposed kernel (\ref{fXC}) 
 should capture the main properties in the multi-orbital system in the case of single-electron per site 
and in majority of cases of several-electrons per site (when J is small comparing to U) .
Most general case of large J's, however, requires additional studies on the form of the XC kernel.
Moreover, the frequency dependence of the corresponding kernel at large frequencies is the same as the exact dependence of the intra- and inter-orbital electron self-energy (Eq.\ref{SigmaR}), thereby supporting the form of proposed XC kernel in Eq.(\ref{fXC}).

The most essential features of the proposed XC kernel are spatial locality, proportionality to the local Coulomb repulsion,
 and the oscillating in time interaction of the electrons with opposite spin, $\sim \exp (\pm iBt)$,
where the frequency of oscillations B is proportional to the hopping (kinetic) energy. 

While further tests of the validity of the methodology presented here for larger scale systems, both in- and out-of-equilibrium is necessary, is on-going, the frequency dependence of the XC kernel in Eq.~(\ref{fXC}) is sufficient to describe the main features of single-orbital correlated systems. We expect the present approach to open the possibility of describing strongly correlated materials within the standard TDDFT framework, extending it beyond the metallic and semiconductor structures with plasmons, excitons and other excitations (see, e.g., Refs.~\onlinecite{30,31}). 

{\it Acknowledgements}.--We thank Lyman Baker for critical reading of the manuscript
and DOE for a partial support under grant DOE-DE-FG02-07ER46354.

\end{document}